# Intelligence Stratum for IoT. Architecture Requirements and Functions

Edgar Ramos
*Ericsson Research, Finland*
edgar.ramos@ericsson.com

Roberto Morabito
*Ericsson Research, Finland*
roberto.morabito@ericsson.com

*Abstract*—The use of Artificial Intelligence (AI) is becoming increasingly pervasive and relevant in many different application areas. Researchers are putting a considerable effort to take full advantage of the power of AI, while trying to overcome the technical challenges that are intrinsically linked to almost any domain area of application, such as the Internet of Things (IoT). One of the biggest problems related to the use of AI in IoT is related to the difficulty of coping with the wide variety of protocols and software technologies used, as well as with the heterogeneity of the hardware resources consuming the AI. The scattered IoT landscape accentuates the limitations on interoperability, especially visible in the deployment of AI, affecting the seamless AI life-cycle management as well. In this paper, it is discussed how to enable AI distribution in IoT by introducing a layered intelligence architecture that aims to face the undertaken challenges taking into account the special requirements of nowadays IoT networks. It describes the main characteristics of the new paradigm architecture, highlighting what are the implications of its adoption from use cases perspective and their requirements. Finally, a set of open technical and research challenges are enumerated to reach the full potential of the intelligence distribution's vision.

*Index Terms*—Artificial Intelligence, IoT, Distribution

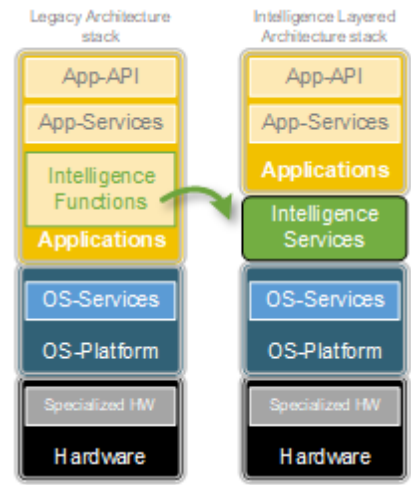

Fig. 1. Intelligent services decoupling.

## I. INTRODUCTION

Intelligence and smartness are considered the top value of the Internet of Things (IoT) [1]. The main reason for connecting things is that they enable the possibility of automation, especially the type of automation that resembles intelligence. There is an expected smart behavior from IoT systems or applications even if the devices are not permanently connected. The nature of such intelligence is as varied as the number of applications in IoT. Thereby, producing intelligent functions for IoT requires good knowledge in applications domains, the deployment of the system, and the involved devices.

A framework proposal was presented in [2] to address the growing complexity of intelligence application development for IoT. The framework introduces an abstraction of intelligence services to an intelligence layer that serves the applications requesting intelligence. The main idea is to decouple the implementation of intelligence functions from the actual application code, thus enabling the possibility of keeping independent Life Cycle Management (LCM) of each intelligent function in one device. The framework explores the possibility of provisioning intelligence functions directly to a device. In contrast, commercially available approaches focus on provisioning Application Programming Interfaces (APIs) for applications to access intelligent services from a cloud network.

In this article, first are identified the use cases and use cases actors of the intelligence layer [2]. Next, the requirements of the intelligence layer are analysed based on the use cases description and the interaction with the use-case actors. Additionally, it is applied to the IoT context and high-level architectural functions are formulated to address the introduced requirements. The use cases are composed following the methodology, notation and concepts specified for the Unified Modeling Language (UML) by the Object Management Group [3]. A UML use case is meant to capture useful systems behavior by specifying the functionality performed by one or more observed subjects to which the use case applies in collaboration with one or more entities or actors.

The rest of the paper is organized as follows. Section II introduces the decoupling concept of the intelligence services from the application layer. Section III describes the special context that IoT introduces when considering distributing AI to devices. Following section analyzes the requirements from the different actors interacting with the intelligence layer's use cases. Section V presents and describes a high-level architecture proposal of the intelligence layer based on the functions decomposition identified to address the use case requirements. Section VI lists additional aspects to be investigated. Finally, conclusions are drawn in the last section.

## II. INTELLIGENT SERVICES DECOUPLING

Intelligent applications and intelligent services are used as interchangeable terms in many situations. In this article it is made a distinction between both. They are distinctive of each other because their end-aims are different. An intelligent application makes use of one or several intelligent services that allows to fulfill its task, which is closely related to a use case. For example, a word processing application may use a voice recognition service to fulfil the task of generating documents for a user. In contrast, the goal of the speech recognition intelligent service is to digitize the oral input of natural language in a format that can be understood and manipulated by other software component. In this example, the software component utilizing the speech recognition service is the word processing application.

Multiple intelligent services can be integrated and serve one application. They may use the output, or part of the output of one, or several other services to produce a new result. This is referred to as service composition. The service composition may be highly integrated in the implementation code of the application, or loosely coupled in a modular fashion. The intelligent services concept is more evident and visible when the applications architecture is modular.

The intelligent services development is different than the development of traditional applications. The programming languages, the development process, the required support data and the acceleration support differ on their requirements and final result. If the intelligent services are directly integrated into the applications, all these factors complicate the life cycle management of the applications and makes their evolution more difficult [4]. Figure 1 exemplifies a device's architecture stack with a layered and decoupled intelligence, in contrast to the legacy application coupled intelligence.

It is possible to find highly decoupled intelligent services from applications in specific domains such as object recognition [5]. Applications utilize REST APIs to access resources in remote locations. The remote resources process input data from the application and reply with information processed by their intelligent algorithms in a format understandable by applications following the API definition. Although there is some degree of decoupling of the services from the application, the main reason for the service break-up resides on the need of offloading computational processing and reduction of data transmission for training purposes. In consequence, the intelligence services are mostly running in remote servers and not directly in the devices where the applications are executed. At the same time, the data required to train the models are centrally collected, and in turn facilitating the training process of the models used by the intelligent services [6].

## III. INTELLIGENCE SERVICES FOR IOT

IoT is a heterogeneous and diverse environment in terms of topologies, protocols, use cases, business processes and deployed technologies. The applications targeted for IoT are also diverse and varied in their purpose, requirements and processing characteristics. Intelligence services serving those applications face similar challenges and require high degree of versatility and adaptability. In some cases, the intelligent services may need to be executed in relatively constrained environments with limitations in memory and processing power, or their processing is offloaded to remote execution environments such as a cloud. The consolidation or aggregation of data may also further restrict how the intelligence services are being handled, including their inputs and outputs.

There are several aspects of intelligence distribution to consider when understanding what Distribution of Artificial Intelligence (D-AI) involves in practice, especially in the IoT context. In first place, there is an intelligence functional distribution. The intelligent services can be decomposed in smaller components or functions with concatenated outputs that feed another function – basically, a modular intelligence approach distributing the intelligence service across modules. A second aspect of D-AI is the inference execution distribution. The intelligence service inference can be executed in different domains according to the availability, needs and requirements of the use case. This aspect is especially critical in the IoT case with their heterogeneity of deployments. The domains range from a processor unit or specialized hardware, to a device, gateway or edge node, or a centralized data center. The intelligence services execution domain can be adapted in accordance to the real-time conditions experienced when the service is required. The inference execution should not be confused with the training execution distribution, even when they relate to the same domain's aspect. The training of models is, in some cases, a very intensive CPU and memory hungry procedure. In consequence, the parallelization of the training process is today a common practice used to provide shorter leading times and, in some cases, also for privacy reasons [7]. A fourth aspect is the provisioning of models or intelligence functions to their execution environments (either for training or inference). Meanwhile the previous two aspects are focused in the execution, the provisioning aspect focuses in how to pack the intelligence (scripts, exchange formats, virtual machines, containers, etc.) and expose the functionality to applications making use of the local resources, following local and remote set policies and enabling the intelligence functions life cycle management. A final aspect to consider is the agent functional distribution. Rational agents may interact between themselves following their purpose in a individual and isolated fashion, but also may follow cooperative behaviors (for example in a hierarchical model, or a swarm model), or even competitive approaches (adversarial models or game theory based optimizations) [8]. The way how the intelligent agents, which may make use of one or several intelligence services, interact with other intelligent agents is also relevant when looking at D-AI for IoT.

## IV. INTELLIGENCE LAYER USE CASE FOR IOT D-AI

The intelligence layer concept introduced in [2] provides services to several use cases' actors that interact with such a layer. Each of the actors has their own set of requirements and interfacing with the intelligence layer of devices aiming for optimal and sustainable operation (scalable, secure, resilient, etc.).

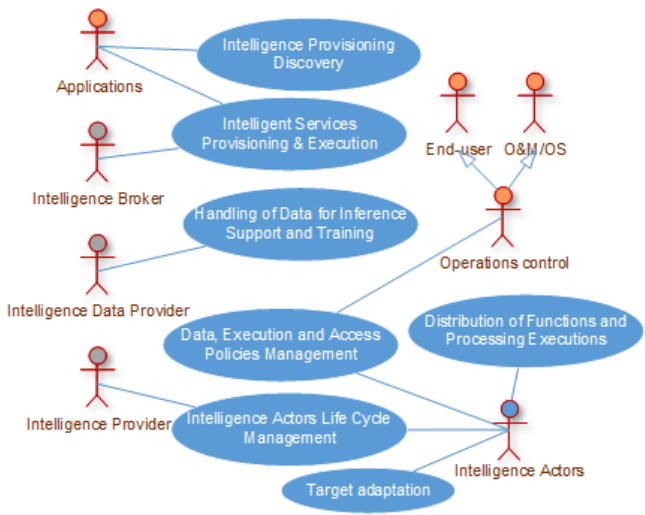

Fig. 2. Use Case Diagram for the Intelligence Layer Services

Figure 2 depicts the UML use case diagram for the intelligence layer reflecting the services required of each of the actors that interact with it. The actors correspond to the main entities identified as ecosystem parties from the distributed machine intelligence ecosystem in [2]. The LCM Intelligence Broker, or simply *Intelligence Broker* in the context of this article, corresponds to the Intelligence Market from the Machine Intelligence Ecosystem. This broker acts as mediator between the intelligence providers and the intelligent devices. It provides the necessary capabilities and support to interactions between the related actors. From the device perspective, the intelligence layer provides a single point of contact with the intelligence broker. The devices utilize their secure connection with the broker to request intelligent services that later are served by a selection of registered intelligence providers to such broker.

The *applications* are the "user" entities of the intelligence services. Before they can use the services, it is necessary to discover where to find them, if they are not already available in the application's device. In this situation, the broker is used to navigate, filter and query – from the registered providers – which intelligent services offered are suitable for the needs of the device's application. Once the suitable service is selected, the intelligence provider service can be provisioned to the device. The service realization may be of diverse nature, for example, a virtual execution environment that can directly serve the application, or a model that is executed in a AI-runtime environment and exposed to the application as a service, or even a call re-direction to a remote executed service that implements the intelligent function.

Once the intelligent service has been provisioned and is ready to be used by applications in the device, it may need to locally be trained, and/or may use some additional data for inference. The *Intelligence Data Providers* utilize either the intelligence market services or communication channels pre-established by the intelligence services provisioned to deliver the required data. The *Intelligence Data Provider* might even be the same entity than the Intelligence Provider, but in this role the focus is on provisioning data instead of intelligence models. The Intelligence Layer must support the data provisioning by supplying storage, and processing services, and mapping the data for inference and training processes, including access control, confidentiality and integrity protection.

Users of devices could be of different kind. Users that own and take advantage of the device capabilities can be identified as direct users straightforward. The ambiguity arises when the users do not necessarily own the device and they may benefit from one function or aspect of the device, or their services are shared with other users, or the real user of the device is another subsystem and not necessarily a human being. This is the case of automated control systems, such as the assembling robot of a factory, or the air-conditioning thermostat of an intelligent house. For this kind of systems, there may have been a human acting as machine operator, as well as humans that benefit from some of the device operations (for example the people in the room being air-conditioned). However, they may not own the device and therefore is difficult to define who the end-user of the device is. Nevertheless, the device is in need of configuration, monitoring and operation of some kind. We could then see this as the Operation and Maintenance (O&M) or Operative System (OS) functions, that many devices include in their systems. These functions effectively control several aspects of the device and they provide interfaces with the relevant entities or "users" to steer the operation of devices. In conclusion, the *operation control* can be modeled as an actor generalization of the roles that a direct user and the O&M/OS systems have with respect to managing devices. The *Operation Control* actor requires access to the intelligence layer to configure policies related to data handling, execution and in many cases accessibility of the devices resources and functionality.

The Intelligent Service (*Intelligence Actors*) is itself an actor in the use case context. The services may be simple functions, named Atomic Intelligent Services (AIS), or Fine-Grained Intelligent Services (FGIS) which are more complex functions composed in many cases by multiple AIS together. The later may even be combined with other FGIS or additional AIS. The result is composition of *Intelligence Actors* that are orchestrated with their inputs and outputs concatenated and serve one or several applications with their intelligence services. The life cycle management of intelligence would apply not only to the intelligence actors but also to the orchestration or composition of such actors, since the *intelligence provider* may use different combination of actor concatenations without updating the individual actors. For example, if the *intelligence provider* to improve the processing efficiency then modifies the order how some AIS are applied on the input data.

The intelligence layer must manage the updating of the *intelligence actors* that have been provisioned and control the dependencies created by the service composition. The updating and modification of the actors have to be verified and the changes authorized by the management policies. Also, the actor's execution environment must be considered and, in some cases, the performance required by certain applications cannot be satisfied by the local execution hardware or by the connectivity characteristics linking to remote execution environments. The conditions may vary according to load, mobility state, concurrency, interference, and others, implying that the intelligence layer may need to redeploy the execution of actors in the best domains available for the specific function that satisfies the performance metrics. In turn, this function is related to the management of data processing according to the policies that the operation control use case actor has provided, because there may be restrictions of data leaving a device or about being processed elsewhere to the device domain.

### A. Requirements from Use Cases Actors

The use case actors' requirements for the intelligence layer can be determined once all the relevant use case actors and their interactions have been identified. In Table I a list of requirements has been compiled for the services that the intelligence layer has to provide to all the use case actors. The requirements have been proposed in line to the following principles for D-AI value generation enablers from [2].

- Technology providers
  - (Use) Proprietary and open models from generic or specialized modules,
  - (operate on) local, remote , public and private infrastructure and
  - (employ) proprietary, 3rd party or local data sources.
- Interoperability space
  - Standardized intelligence modelling and exposure to applications.
  - Harmonized intelligence life cycle management and provisioning discovery.
- Devices and Execution
  - Open intelligence services platform.
  - Services or support to any hardware type.
- Intelligence Functions
  - Mostly composition of generic solutions.
  - Peer-to-peer, multi-homing, stand-alone and client-server applications.

## V. INTELLIGENCE STRATUM FUNCTIONS

In this section, it is described what are functions and minimal requirements that the layered intelligence architecture must embed in order to fully enable the envisioned ecosystem. Figure 3 is referred to provide an high-level view. The term "compass" is used to convey the idea that the intelligence layer is the core element of this new intelligence paradigm, and to emphasize the fact that its definition impacts almost every single aspect of the whole ecosystem, both from technical and business perspective.

High level functions can already be devised to address the requirements to the intelligence layer services from Table I. The functions are naturally defined based on their interactions with the other actors and therefore realizing the functionality that the interfaces exposes.

### A. Intelligence Layer Compass

The intelligence layer is characterized by a set of functional components forming its core and four distinct interfaces (Northbound, Southbound, Eastbound, Westbound at Figure 3).

Northbound and Southbound interfaces resemble the typical structure of regular IoT platforms. The main purpose of the Northbound module is to interface with overlying applications as well as user/system management engines. The Southbound module acts as a sort of device manager, meaning that it handles, selects and executes all hardware and supplementary libraries needed for ensuring the most suitable runtime system for all the Intelligence Layer components running on top of a given hardware platform. Furthermore, while controlling the execution, it ensures that application's QoS and performance requirements are satisfied. Westbound and Eastbound interfaces have been introduced along with the intelligence services decoupling concept to fully enable this new intelligence paradigm vision with its entire ecosystem. The Westbound module is meant to take care of all the Life Cycle management aspects, as well as provisioning, update and discovery of intelligence. The Eastbound interface, in contrast, is responsible for ensuring the remote and distributed execution of intelligence services. From this perspective, the eastbound interface owns the key role of guaranteeing an efficient coupling with associated intelligence layers. This latter task is not trivial as it must be ensured that multiple intelligence layers can be associated according to specific topology (e.g., master-slave, peer-to-peer, etc.). Furthermore, such association requires trust management services to be executed to discriminate legit from tampered instances and prevent any potential attack. In this respect, securing the entire data pipeline between the associated intelligence layers becomes also a key task that such interface needs to cover.

The **Northbound interface (NB)** bridges the functionality provided by the intelligence layer with the overlying application running on top of it. It enables key processes such as automatic setup and discovery of intelligent functions dependencies, updates and redeployment of intelligent services, access to intelligent functions chain, execution and inference policies, etc. Consequently, there are multiple aspects that needs to be handled by such interface and most of them relate on the setup, management, and release of the APIs existing between the application and intelligence layer. Additionally, there are a set of policies and customization rules that must be defined to handle the data flowing among application and other interfaces.

TABLE I
SERVICES REQUIRED FROM INTELLIGENCE LAYER

| **Actor: Intelligence Broker.** | **Actor: Intelligence Data Provider.** |
|---|---|
| **Use case:** Intelligence Provision Discovery.<br><br>- Trusted and secure communication to and from Intelligence Brokers.<br>- Intelligence query and handshake initiation handling for intelligence on-boarding and updates.<br>- Multiple brokers accessibility. | **Use case:** Training and Inference Support Data Provisioning.<br><br>- Interoperable and secure data ingestion from third party providers to devices.<br>- Verification and certification of intelligence data origins.<br>- Enforcing of device system policies for outbound processed data.<br>- Data providers policies enforcement for data treatment in the device. |
| **Actor: Intelligence Provider.** | **Actor: Applications.** |
| **Use case:** Intelligence Actors Life Cycle Management<br><br>- Interoperable on-boarding, setup, updating, execution and decommissioning of intelligent functions (actors) in a device.<br>- Handling of intelligent actors from trained or untrained models using a standard or open intelligence exchange format.<br>- Training tools & methods support for untrained models.<br>- Intelligent services composition (chaining atomic intelligent services together).<br>- Verification and certification of intelligence functions origins.<br>- Device capabilities reporting. | **Use case:** Intelligence Provision Discovery<br><br>- Automatic setup and discovery of intelligent functions dependencies from applications.<br>- Updates and redeployment of intelligent services.<br><br>**Use case:** Intelligence Services Provisioning and Execution<br><br>- Access to intelligent functions chain according to access and execution policies.<br>- Execution and inference policies enforcing of application specific data. |
| **Actor: Intelligent Actors (or intelligent services).** | **Actor: Control Operators (end user or O&M system).** |
| **Use case:** Intelligence Actors Life Cycle Management<br><br>- Versioning and re-deployments of actors and supporting of the whole actors life cycle management.<br><br>**Use case:** Functions and Processing Execution Distribution<br>- Composition of intelligence actors in chains according to access and execution policies.<br>- Intelligence actor's specific data, execution and inference policies.<br>- Execution orchestration in multiple domains to maximize performance.<br><br>**Use case:** Target Adaptation<br><br>- Interfacing between the device hardware environment and higher level functions, specially for acceleration purposes | **Use case:** Data Execution and Access Policies Management<br><br>- Configuration and customization of execution, training and inference polices of intelligence actor's specific data in applications, execution domains and intelligent services.<br>- Monitoring services and a rule engine to match performance targets or contingency actions during operation.<br>- End-user agreements configuration for provisioned intelligent services |

Configurable APIs are required to enable a flexible and seamlessly interaction between the intelligence layer and applications. From this perspective, hypermedia can be considered a suitable technology to achieve such flexibility. Although hypermedia is still not used to its fullest extent in the IoT context, lately there have been several attempts to make sure that web technologies can be also used in different deployment environments [9]. The advantage brought by the potential use of web applications (e.g. hyperlinks, forms, web media types) strives by the possibility of making the system dynamically adaptable to the changes without undermining usability and compatibility between the two interfaced layers. Furthermore, the NB must embed mechanism allowing the application to semantically discover how to find and use the APIs when, for example, intelligence layer and a specific application are associated for the first time. The NB characteristics enable a new way of writing intelligent applications, since developers may not code the *intelligence* directly into the application, but rather make sure that application's APIs can fully interact with the underlying layer and only fetching the intelligence services from it.

Additionally, the NB provides support to the operation control for configuring policies and customization rules. They are needed for managing how data flows through the intelligence layer, by considering also supplementary aspects including data confidentiality, privacy, data storage and data access, data re-usability, data provenance, data ownership, etc. Furthermore, the policies govern the interactions of intelligence with the resources available to it, such as, management of connectivity, or the performance metrics to achieve a determined quality of service. These policies can be handled by a rule engine that dynamically adjust the resource utilization and matching intelligence model execution according to the applicable policy that might restrict the execution domain or constrain how much processing load, memory or power can be spared. In order to validate the consistency of enforcing policies, the actor management embeds watchdogs that constantly verify whether there is compliance with respect to the established rules. In reference to the *compass*, the Intelligence Services Exposure is the functional entity exposing the APIs that allows the association between the intelligent functions (intelligence actors in the intelligence layer context) and the intelligent applications, meanwhile the actors management functions are performed by the Actors Engine. The Intelligence Actors life-cycle and association to applications is handled by the Actor Repository, while the policies are handled by the Management Engine, which has specific functions related to monitoring, data and intelligence actors' policies.

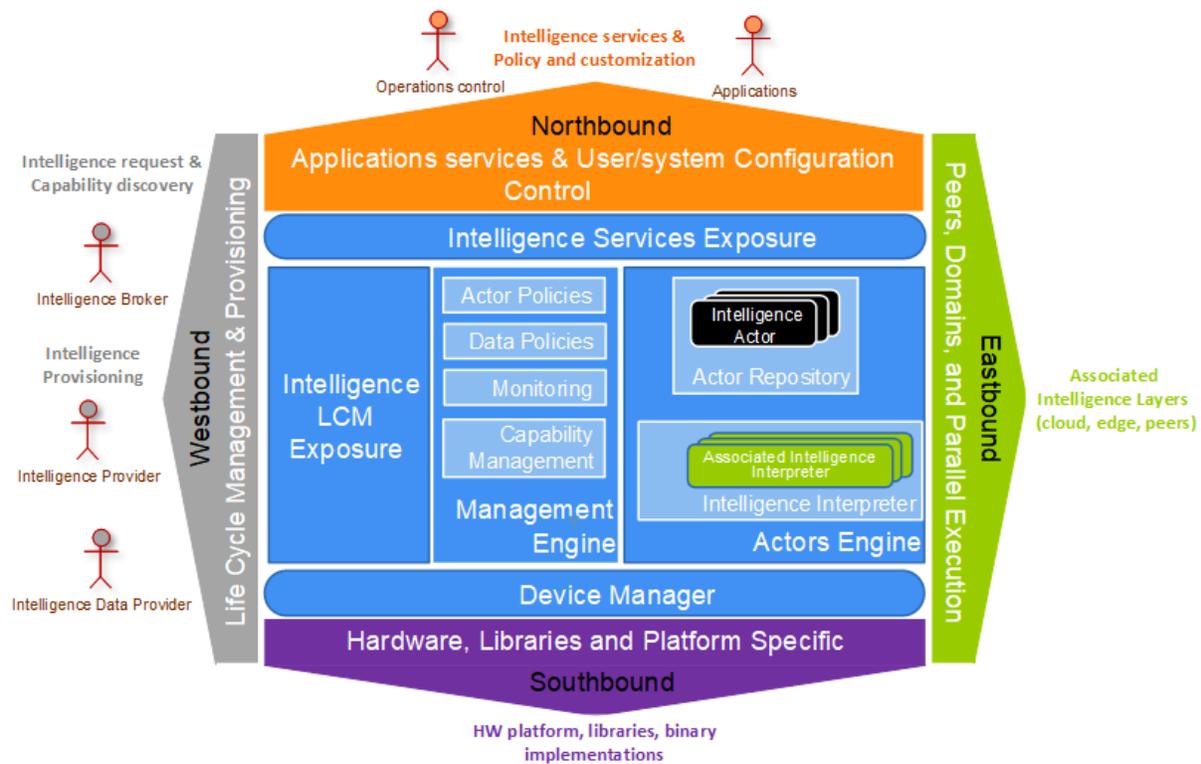

Fig. 3. Intelligence Layer Compass

The **Westbound interface (WB)** plays a key role to fully realize the intelligence layer-based ecosystem and make possible carry on most of the envisioned use cases. The specialization of the WB relies on maintaining the life-cycle management of the intelligence services as well as managing the services provisioning, by enabling the interactions with intelligence providers and intelligence data providers as external entities. The Intelligence LCM Exposure is the intelligence layer component that realizes the WB interface providing access to the internal components of the intelligence layer. It also ensures that all the LCM management related tasks are executed, as well as guarantees trusted and secure interactions with intelligence providers and intelligence brokers (Figure 3).

It varies from case to case to what extent intelligence data provider and intelligence provider are the same entity. However, there are multiple cases in which these two entities are different and therefore each of them needs a separate API for interacting with the actors' management. The intelligence data provider API must be designed in a way to ensure data confidentiality – meaning that data are used only for the purpose for which they are received –, but also to make sure that the actors management can easily recognize the correlation/correspondence to the AI model for which the received data can be used. In this respect, semantic-based solutions to describe data can be useful for simplifying and make transparent this association, for example, labeling as weather data a subset of temperatures values. From the technical point of view, the WB APIs may be activated bidirectionally and grants real-time or cached access both to the actors engine and/or intelligence data provider. The API between intelligence provider and intelligence layer can be realized differently depending on whether the provider delivers the intelligence by direct interaction with the intelligence layer or, alternatively, by a mediator intelligence broker that fetches the intelligence services through dedicated market or similar entities. There are clearly different trust and security implications in the two foreseen scenarios, which also depends on the relation established between the parties. For each case, additional policies need to be defined. In any case, it is expected that a direct involvement is more suitable for industrial use-cases where dedicated and stronger agreements between providers and users are expected. Differently, for general purpose and consumer cases is likely that a market-based approach is preferred. In both cases though, there are several design and implementation aspects that need to be considered. For example, it is important to define an interoperable, extensible, and open methodology that allows applications to easily benefit from using a wide set of AI models/frameworks, selected according of the different phases of the intelligent application development (e.g. training or inference) and of the additional system requirements (e.g. from network architecture perspective) [10]. Integrity check and certification of the services provided represent a key aspect as well. To this extent, standardized mechanisms, similar as IETF (Internet Engineering Task Force) software updates for IoT [11] may be used to satisfy such requirements.

Returning to the role of the intelligence broker in the discovery of intelligent services, a metadata-based approach is advisable. Additional semantic annotations can be defined to describe the appropriate level of detail needed to expose the services. The handshake initiation for intelligence onboarding and updates must be defined effectively, to model the interaction between providers, broker and intelligence layer. In fact, additional mechanisms must be defined mainly for mapping the supplier policies to the device policies (e.g. data policies and actor policies). It is also important to understand, how to effectively reuse existing protocols and networking technologies. It is needed to optimize the intelligence handshake taking in account the policies mechanisms and other additional aspects also, such as, devices heterogeneity and their capabilities.

The **Southbound interface (SB)** bridges intelligence layer and the underlying hardware platform. In order to effectively tailor the intelligence services to the hardware and software capabilities of the platform, such interface receive and process a relevant amount of information as it interacts – mainly through the Device Manager – with several intelligence layer functional blocks (i.e. management engine, actors engine). The SB is responsible for transferring information to the device manager about hardware discovery, software availability and resource availability making in practice the intelligence layer hardware and software agnostic. From hardware discovery perspective, the intelligence layer must be able to determine whether is interacting, for example, with a Graphics Processing Unit (GPU) rather than a Tensor Processing Unit (TPU), a Single-Board Computer (SBC), or a very constrained device. This awareness of the available hardware resource of the device limits what software can be executed in the device itself. In terms of software capability, the interaction device-to-device manager is mainly based on understanding whether the device embeds already all the needed software components to support the requested intelligence service. It is worth clarifying that in this process also the WB can be also involved, as the intelligence service may require an updated or newer AI algorithm that is not stored or cached already in the device and therefore needs to be requested and provisioned from an intelligence provider. In both cases of software and hardware capabilities discovery, an hypermedia-based approach can be used for enabling it. However, taking into consideration the IoT context, we also envision the use of Constrained RESTful Application Language (CoRAL) [12] as one possibility of enhancing discovery capabilities also in the device-constrained space, and of the Lightweight M2M protocol (LWM2M) [13] for reporting hardware and software capabilities.

To maintain a real-time overview of all the available device resources and react to changes or variations of the environment or resource utilization, dedicated HW and SW monitoring capabilities – hosted and provided by the management engine – are essential, for example, to act according to the type of connectivity available, or power supply state – corresponding to the management policies provided.

At last, the intelligence layer delegates on the **Eastbound interface (EB)** to perform all the procedures connected to the distributed execution of intelligence services. In this context, distribution of execution means enabling a seamless interaction with other intelligence layer peers or complementary intelligence-based domains, but also the issuing of parallel execution among devices or the off-loading of intelligence service execution to other peers. This function is offloaded to the Intelligence Interpreter, who has to interact with the Device Manager and the Capability Management to decide on the best execution strategy for each of the intelligence actors of an intelligent service.

The execution of intelligence services across multiple domains, as well as the association between several intelligence layers brings additional challenges that need to be addressed. To this extent, the role played by the EB is important to ensure seamless association and on-boarding mechanisms, continuous trust management, secure data flow and data provenance, and actor re-deployment (in the event that intelligence services are offloaded). Referring to the case of association between several intelligence layers, additional aspects are worth to be clarified further respect the role of the EB. The first one relates to the fact that in the association process, the EB shall be capable to negotiate what kind of distribution model must be established (e.g. master-slave or peer-to-peer) with the associated intelligence layer(s). In this process, management engine and device manager are also involved to ensure that the association becomes feasible both from hardware and software perspective. Then, the EB must be designed to support that the *associated intelligence interpreter* receives all the necessary information for exploiting the coupled intelligence. Finally, the EB need to support the orchestration of the (distributed) actor re-deployment. In this respect, existing solutions such as [14] are useful, if considering additional several conceptual and technological extensions (e.g. owning the characteristic of being agnostic respect both applications' programming language and runtime execution environments).

## VI. OPEN TECHNICAL AND RESEARCH CHALLENGES

The concept introduced in this paper defines a novel way of exploiting the power of AI in a wide set scenarios with special emphasis in IoT. Nonetheless, several technical areas deserve further investigation to fulfill the views expressed and progress towards the implementation of a first deployment. First, the main goal must be to demonstrate the feasibility of decoupling the intelligence from the application understanding the potential benefits deriving by the introduction of the intelligence layer paradigm both from deployment and business perspective. From a purely technical point of view, there are several implementation practices that are worthwhile to consider when building a prototype. In particular:

- Provisioning of AI models and intelligence service composition in a layered intelligence architecture.

- Determine the choices of protocols to use by and among the different intelligence layer functions and interfaces. Also, select between the different architectural styles (e.g. REST vs. publish-subscribe) and interaction paradigms (e.g. request-reply, callback, or notification based) according to the different requirements.
- Define Intelligence Services API discovery with semantic descriptors.
- Orchestrate Intelligent Actors in and between devices/domains.
- Set up protected data pipelines (separation of concerns, shared data access, semantic based security and privacy, policy distribution and contextualization).
- Further explore trust establishment and propagation to data management in a device (e.g. through ledgering technologies).
- Evaluate the suitability of widely-used OS (e.g. Linux or Android) as underlying software platform for a full realization of our concept – also hardware portability is considered crucial.
- Identify what monitoring and capability management mechanisms for improving platform and device efficiency are needed.

Linked to the above listed challenges, there are also a set of non-extensive open research questions that are listed below.

- How to do secure, private and protected interoperable service composition?
- What technologies can be used to harmonize and make extensible the APIs enabling the interaction between applications and intelligence layer?
- How to evaluate the balance between complexity, performance and efficiency of generic IoT AI models considering device available resources?
- How to find the best suitable policy handling strategies for data and intelligence execution?
- What are additional enablers required for the distribution ecosystem friction-less operation?
- What are the mechanisms to handle the execution distribution of intelligence services (both from hardware and software perspective)?

It is expected that the above listed research questions and technical challenges attract the interest both from researchers and developers in working in the development of all these different aspects.

## VII. Conclusion

In this paper, it is introduced and described a novel way of realizing how to deploy AI, especially in the IoT domain. It is based on the idea that intelligence services decoupling from the applications enables a more flexible and open intelligence life-cycle management, and interoperability with several additional benefits than coupled approaches. It is also presented what are the technological-based grounds that can bring forward the new paradigm, by highlighting also how new use cases and opportunities are created from its technological development. They have been turned on requirements that are addressed for specific functions described in high-level and mapped to components of the intelligence layer serving applications in a device. In particular focus was the interaction between the intelligence layer and external actors that utilize its services. Also, in focus was its applicability in the specific context of IoT, as one possible extreme case scenario in which intelligence services must be handled effectively, by taking into account the constraints due to the hardware, software, and networking heterogeneity.

Finally, a list of critical technical challenges and promising research areas have been drawn to provide specific guidelines for future works to a wider research community. As a next step, it is encouraged further developments towards the realization of the intelligence service decoupling, studying its practical feasibility and the ecosystem impact generated by it.